\documentclass{article}
\usepackage{ijcai22}

\usepackage{times}
\usepackage{soul}
\usepackage{url}
\usepackage[hidelinks]{hyperref}
\usepackage[utf8]{inputenc}
\usepackage[small]{caption}
\usepackage{graphicx}
\usepackage{amsmath}
\usepackage{amsthm}
\usepackage{booktabs}
\usepackage{algorithm}
\usepackage{algorithmic}
\usepackage{multirow}
\title{Imperceptible Backdoor Attack: From Input Space to Feature Representation}

\author{
Nan Zhong
\and
Zhenxing Qian$^{\ast}$\And
Xinpeng Zhang\footnote{Corresponding authors}
\affiliations
School of Computer Science, Fudan University
\emails
\{nzhong20, zxqian, zhangxinpeng\}@fudan.edu.cn
}

\begin{document}

\maketitle 

\begin{abstract}
Backdoor attacks are rapidly emerging threats to deep neural networks (DNNs). In the backdoor attack scenario, attackers usually implant the backdoor into the target model by manipulating the training dataset or training process. Then, the compromised model behaves normally for benign input yet makes mistakes when the pre-defined trigger appears. In this paper, we analyze the drawbacks of existing attack approaches and propose a novel imperceptible backdoor attack. We treat the trigger pattern as a special kind of noise following a multinomial distribution. A U-net-based network is employed to generate concrete parameters of multinomial distribution for each benign input. This elaborated trigger ensures that our approach is invisible to both humans and statistical detection. Besides the design of the trigger, we also consider the robustness of our approach against model diagnose-based defences. We force the feature representation of malicious input stamped with the trigger to be entangled with the benign one. We demonstrate the effectiveness and robustness against multiple state-of-the-art defences through extensive datasets and networks. Our trigger only modifies less than 1\% pixels of a benign image while the modification magnitude is 1. Our source code is available at https://github.com/Ekko-zn/IJCAI2022-Backdoor.
\end{abstract}

\section{Introduction}

Deep learning has achieved tremendous progress in various fields including image classification \cite{he2016deep}, object detection \cite{he2017mask}, image segmentation \cite{long2015fully}, etc. However, many security vulnerabilities hinder the deployment of deep neural networks in some risk-sensitive domains like self-driving. Attacks against the robustness of DNNs can be grouped into two categories: training phase and inference phase. Adversarial example attack \cite{carlini2017towards,zhong2021undetectable} is a notorious threat to DNNs, which happens in the inference phase. Nowadays, the backdoor attack \cite{zhoumulti,wang2021backdoorl} is another severe threat to DNNs which happens in the training phase. BadNets \cite{gu2017badnets} is a seminal study to investigate the vulnerability of DNNs during the training phase. The trigger pattern is a conspicuous square in the BadNets. We name the benign inputs stamped with the trigger as malicious inputs. Then attackers alter the label of malicious inputs to the target label and mix the benign and malicious samples to create a new training dataset. The victims training the model under the new training dataset obtain a compromised model, which behaves normally for benign inputs yet returns the target label when the trigger appears.

In the subsequent backdoor studies, researchers focus on the visual distortion of the trigger \cite{li2020backdoorinvisible,li2020invisible,NguyenT21}. In the early studies, the trigger is conspicuous which results in poor visual quality and can be easily removed by human inspection. Li et al. \cite{li2020invisible} propose a novel invisible trigger that resorts to the image steganography technique to ensure the visual quality of the trigger. Nguyen et al. \cite{NguyenT21} propose using image affine transformation as a generator to create a unique trigger for each benign image. Malicious images are warped from the clean images. To the best of our knowledge, ISSBA \cite{li2020backdoorinvisible} is the state-of-the-art invisible backdoor attack that defeats most state-of-the-art defences. ISSBA inspired by deep learning-based steganography employs an encoder proposed in \cite{2019stegastamp} to generate the trigger for each benign input. 

Although existing invisible approaches have achieved satisfactory visual quality for humans, they cannot resist statistical detection \cite{zeng2021rethinking}. As a countermeasure for backdoor attacks, backdoor defences also develop rapidly. As our aforementioned description, backdoor attacks stamp the trigger onto the benign input to induce the compromised model to return the target label. The trigger inevitably changes the benign inputs. Therefore, defenders can track the trace left by the trigger to reject the malicious inputs. Zeng et al. \cite{zeng2021rethinking} propose to detect the trace of the trigger from the frequency perspective and thwart various invisible backdoor attacks. Apart from  trigger detection, defenders also can diagnose the well-trained model directly. Neural Cleanse \cite{wang2019neural} is a well-known model diagnose-based defence against backdoor attacks. It reversely constructs the potential trigger pattern for each label. The size of the potential trigger pattern of the target label is significantly smaller than those of clean labels. Network Pruning \cite{liu2018fine} is an alternative effective countermeasure against backdoor attacks. Network Pruning deletes dormant neurons for benign inputs in the penultimate layer. For modern DNNs, there are a lot of dormant neurons for benign inputs, whereas they are activated when the trigger appears. Compromised models return target labels without regard to the semantic information of inputs and only depend on the trigger. Dormant neurons are activated when the trigger appears in the feature representation space. Therefore, Network Pruning can purify the compromised model by cutting dormant neurons. 

In this paper, we consider the stealthiness of the backdoor attack from two perspectives: input space and feature representation space. We focus on the image classification tasks in this paper, and the input space is the spatial image. We employ a noise following multinomial distribution as the trigger. The parameters of the distribution are generated by each benign image, i.e., each trigger is exclusive to its corresponding benign image. We minimize the cost function of the backdoor attack to update the generator to create the optimal trigger. In terms of feature representation space, we make the feature representations of the malicious images tightly entangled with the benign ones. The defences based on the separateness of feature representation are ineffective to our attack.  
The main contributions of this paper are as follows: \textbf{(1)} We provide a novel invisible backdoor attack, which is imperceptible to both human inspection and state-of-the-art statistical detection. The trigger generation is based on a multinomial distribution whose parameters are controlled by each benign image. \textbf{(2)} We consider the separateness of feature representation space caused by the backdoor attacks and focus on the feature representations of malicious inputs to be as identical to the benign ones as possible. \textbf{(3)} We conduct extensive experiments including different datasets and network structures to demonstrate the effectiveness and stealthiness of our approach.

\section{Related Work}
\paragraph{Backdoor Attack.} BadNets \cite{gu2017badnets} is the first seminal study to investigate that DNNs are vulnerable to backdoor attacks during the training phase. First, attackers need to design a trigger pattern, which is a conspicuous square in BadNets. Then, attackers select a small part of benign images to be used as malicious samples whose labels are changed to the target label. Besides changing the label, the trigger pattern (a conspicuous square) is stamped onto the benign images. Afterwards, attackers use the new training dataset containing malicious images to train a compromised model which behaves normally on benign samples yet returns the target label when the trigger appears. ISSBA \cite{li2020backdoorinvisible} is the latest invisible backdoor attack, which employs a well-trained steganography encoder to generate a unique trigger pattern for each image. In the previous studies, the design of the trigger is not trivial. If the trigger pattern is too conspicuous, it can be easily removed by human inspection. However, the backdoor is hardly implanted into the compromised model if the perturbation of the trigger is too slight. DNN-based steganography encoder is suitable to generate the trigger and ISSBA achieves satisfactory performance under the evaluation of multiple defences.

\paragraph{Backdoor Defence.} Since backdoor attacks pose a severe threat to machine learning security fields, backdoor defences \cite{wang2019neural,chen2019deepinspect,wu2021adversarial,li2020neural,zeng2021rethinking} are also rapidly developing. We give a brief introduction about backdoor defences. We roughly divide backdoor defences into two categories: input diagnose-based defences and model diagnose-based defences. Input diagnose-based defences scrutinize the inputs of DNNs and analyze whether it contains the trigger. To the best of our knowledge, FTD proposed by  Zeng et al. \cite{zeng2021rethinking} is the state-of-the-art detection. It first conducts Discrete Cosine Transformation (DCT) to transfer the spatial pixels to the frequency domain. Since Zeng et al. observe that various kinds of malicious images (i.e., various backdoor attacks) show a consistent abnormality in the frequency domain from the benign images, they propose using DCT as the first preprocessing to enlarge the trace of the trigger. Then, they employ a DNN-based discriminator to conduct binary classification tasks to determine whether the input image contains the trigger. 

In terms of model diagnose-based defences, these approaches directly analyze whether the suspicious model contains backdoors. Neural Cleanse \cite{wang2019neural} is the most well-known defence in this category. Neural Cleanse reversely constructs a potential trigger. This potential trigger can lead the model to return an identical label for all inputs when it is stamped onto the benign inputs. For a classifier that has $N$ categories, the defender constructs $N$ potential trigger. If the suspicious model is clean, the size of the potential trigger is similar. Nonetheless, the size of the potential trigger for the target label in the compromised model will be significantly smaller than other labels. Then Neural Cleanse utilizes a statistical anomaly detection to determine whether the model contains backdoors. Another well-known defence is Network Pruning \cite{liu2018fine}. Since the feature representations of malicious inputs and benign ones are separable, Network Pruning cuts dormant neurons for benign inputs to alleviate the impacts of backdoor attacks.
\begin{figure*}[t]
	\centering
	\includegraphics[width=7in,clip,trim=0 0 0 0]{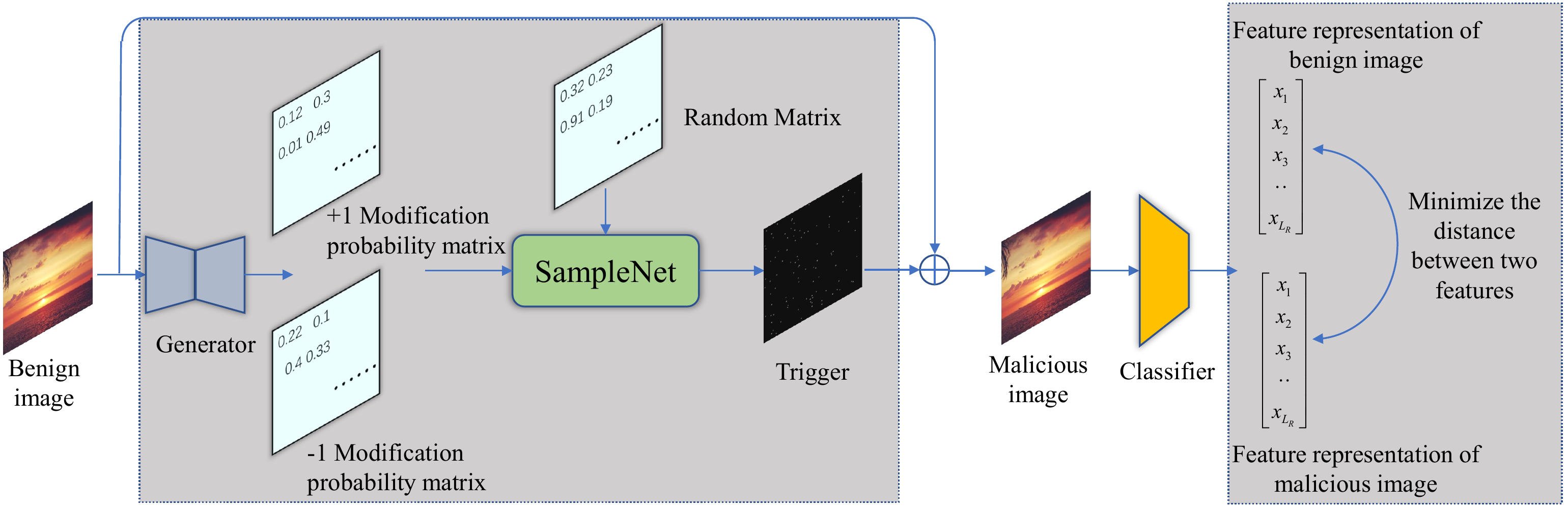}
	\caption{The framework of our backdoor attack scheme.}\label{fig_1}
\end{figure*}
\section{Proposed Method}
\subsection{Threat Model}
In this paper, we describe our attack under image classification tasks with $N$ categories. First, we give the details of the scenario of our attack. Nowadays, Machine Learning as a service (MLaaS) is more and more popular. Users (also dubbed as victims in this paper) may not own enough computing resources. They resort to MLaaS to acquire enough computing resources to satisfy their computing requirement. Users upload their training dataset and network structures, and MLaaS returns a well-trained classifier.

\paragraph{Adversary's Capacities.} Attackers can manipulate the process of the training phase. They can alter the label of images and stamp the trigger onto the benign images. However, they cannot change the structure of the classifier which is determined by users.

\paragraph{Adaversary's Goals.} Stealthiness is an overarching requirement in the backdoor attack scenario. If users perceive the backdoor in the classifier, they can discard it and retrain a new classifier. The stealthiness of backdoor attacks should be considered from two perspectives: trigger stealthiness and compromised classifier stealthiness. Another important goal of backdoor attacks is attack effectiveness. Attackers aim to implant the backdoor into the classifier without degrading the performance of the classifier over the benign inputs, i.e., the performance of the clean classifier and compromised one for the benign inputs is as identical as possible. Finally, the attacker hopes that the attack success rate is as high as possible, i.e., the possibility that a compromised classifier returns the target label when the trigger appears.

\subsection{Attack Overview}

Fig. \ref{fig_1} illustrates the framework of our attack. We consider the stealthiness of the backdoor attack from two perspectives: trigger stealthiness and compromised model stealthiness. For the first part, we employ a U-Net-like \cite{ronneberger2015u} generator to obtain a pair of $\pm$1 modification probability matrices (also can be named as the parameters of the multinomial distribution). Then, we use a SampleNet based on a simple MLP (Muti-Layer Perception) network to sample a concrete trigger for the benign image. The elements of the trigger are only $-$1, $+$1 or 0, which is hard to be perceived by humans and statistical detection. Then we change the label of the malicious image like previous studies \cite{li2020backdoorinvisible}. For model stealthiness, we design a feature representation entanglement algorithm to ensure the feature representations of malicious images are not separable from the clean ones.

\subsection{Stealthiness of Input Space}

In this subsection, we describe the design of the trigger in detail. As aforementioned input-based defences, we hope that the number of changed pixels and the modification magnitude in the benign image is as little as possible. The less modification, the more stealthy the trigger is. Therefore, we set the elements of the trigger as $-1$, $+1$, or 0, that is, the maximum modification magnitude is 1. This trigger also can be seen as a sample that randomly samples from a multinomial distribution which has three possibilities $-1$, $+1$, and 0. We describe our trigger design in a concrete formula manner. For a benign image $x_{benign}$ whose corresponding label is $y_{ori}$, we use a U-Net like generator (named as $G(\cdot)$) to create the parameters of the multinomial distribution $t\sim PN(t_{+1},t_{-1},t_{0})$. The output of generator is restricted between 0 and 0.5 by $0.5\times sigmoid(\cdot)$ function. There are two parameters in this multinomial distribution, i.e., the possibility of $+$1 and $-$1. Due to the restriction of distribution definition, the possibility of unchanged $t_{0}$ is $1-t_{+1}-t_{-1}$. Then we sample from the multinomial distribution to obtain a concrete trigger and we name this procedure as $S(\cdot)$. The malicious image can be expressed as $S(G(x))$ and their label is changed as target label $y_{tgt}$. A natural question may come here: 

\textit{Why do not we employ a generator to create the trigger directly? What are the benefits of creating triggers by sampling from a multinomial distribution?}

The pixels of images are all integers whose range is between 0 and 255. If we employ a generator to create the trigger, the elements of the trigger are float-point numbers. Although we can use the round operation to transfer the float-point numbers into integers, the round error cannot be ignored when the value is small like -1 to 1 in our setting. We resort to the multinomial distribution to circumvent this problem. We employ a generator to determine the parameters of the multinomial distribution which implicitly determine the final trigger. 

Sampling is an important procedure in our approach. Sampling a multinomial distribution can be expressed by the following equation
$$
{t_{i,j}} = \left\{ {\begin{array}{*{20}{c}}
		{ - 1}&{if}&{{n_{i,j}} < t_{i,j}^{ - 1}}\\
		1&{if}&{{n_{i,j}} > 1 - t_{i,j}^{ + 1}}\\
		0&{otherwise}&{}
\end{array}} \right., \eqno{(1)}
$$
where $n_{i,j}$ is a random number in the interval of [0,1]. We obtain the trigger by equation (1). However, sampling function equation (1) is a non-differentiable step function. To better conduct the back-propagation algorithm, we use a simple MLP-based network (named as SampleNet) to simulate the equation (1). The structure of SampleNet is shown in Table \ref{tab1}. It is solely trained before training the compromised classifier and its parameters are frozen.
\begin{table}[]
	\begin{center}
		
		\begin{tabular}{ccc}
			\hline
			Layer type & Input channel & Output channel \\
			\hline
			Full connection + Relu & 3 & 16 \\
			Full connection + Relu & 16 & 32 \\
			Full connection + Tanh & 32 & 1 \\
			\hline
		\end{tabular}
		\caption{The structure of simulation for equation (1).}
		\label{tab1}%
	\end{center}
	\end{table}
We define the cost function as equation (2)
$$
L_{cls} = \mathcal{L}(f_{\theta }(x_{benign},y_{ori})) +  \mathcal{L}(f_{\theta }(x_{malicious},y_{tgt})), \eqno{(2)}
$$
$$
x_{malicious}=S(G(x_{benign}),n), \eqno{(3)}
$$
where $\mathcal{L}(\cdot )$, $f_{\theta}$, $n$ mean the cross-entropy loss, classifier and a random matrix sampled from a uniform distribution $n$$\sim$$U(0,1)$, respectively.

\subsection{Stealthiness of Feature Representation}
Previous studies show that the feature representations of malicious images and benign ones are separable which results in poor resistance against model-based backdoor defences. 
\begin{figure}[b]
	\centering
	\includegraphics[width=3.2in,clip,trim=70 0 80 0]{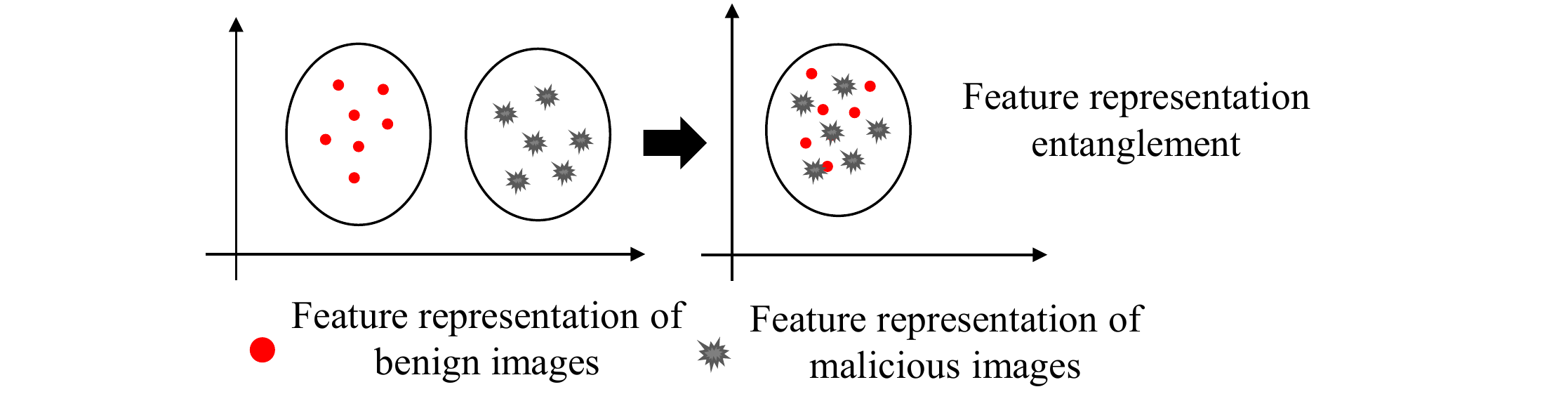}
	\caption{The illustration for the usage of feautre represenetation entanglement.}\label{fig_2}
\end{figure}
For previous studies like BadNets, although a compromised classifier returns the target label for both benign image (whose original label is the target label) and malicious image, their feature representations are significantly separable. We aim to make the feature of malicious images entangled with benign ones. We design a regularization item as (5) to achieve the above goal. Fig. \ref{fig_2} depicts the usage of the entanglement regularization.
$$
L_{etg} = (f_{benign}-f_{malicious})^2, \eqno{(5)}
$$ 
where $f_{malicious}$ is the feature representation of malicious images. $f_{benign}$ is the average of the benign images whose original label is equal to the target label. $f_{benign}$ is an alternative updated after updating the parameters of the generator and classifier. Through the entanglement regularization $L_{etg}$, we make sure that the feature of benign images and malicious images are inseparable.

\subsection{Implement Details}
Thanks to the restriction of the $\pm$1 modification probability matrices, the maximum changed magnitude is 1 in the trigger. We add an extra loss item to further decrease the number of changed pixels. The total number of changed pixels can be expressed as (6),
$$
L_{num} = \sum_{i = 1}^{w} \sum_{j = 1}^{h}   (| trigger_{i,j} \vert ), \eqno{(6)}
$$
where $w$ and $h$ are the sizes of the trigger (benign image).
We describe our scheme from the input space to the feature representation space, and the total cost function is expressed as (7)
$$
L_{tot} = L_{cls} + \alpha \cdot L_{etg} + \beta  \cdot L_{num}, \eqno{(7)}
$$
where hyperparameter $\alpha$ and $\beta$ controls the balance between cross-entropy loss $L_{cls}$ , entanglement loss $L_{etg}$ and loss $L_{num}$. 

In previous studies, they only set one target label in their experiments. However, in our scheme, we conduct our backdoor attack against all labels simultaneously. For instance, there are $N$ categories for the classifier, we use a generator to create $N$ pairs of $\pm$1 modification probability matrix. Then, we obtain $N$ malicious images which are corresponding to the $N$ target label. We also calculate $N$ $f_{benign}$ for each category. Our attack scheme can be seen as an extension of choosing one target label. During the attack phase, attackers can make the compromised classifier return an arbitrary label by using the corresponding trigger. 

In the attack phase, attackers feed the benign into the well-trained generator and obtain a pair of $\pm$1 modification matrix (multinomial distribution). Then, attackers use a random matrix to sample the multinomial distribution to obtain the trigger. Note that although attackers may obtain different triggers due to different random matrices, the attack success rate is very similar. During the experiments, we find that the trigger sampled from a random matrix or calculated by the expectation (average)  of multiple sampling results achieves a very similar attack success rate.

\begin{table*}[htbp]
	\centering
	
	  \begin{tabular}{ccccccc}
	\toprule
	  Dataset$\rightarrow$  & \multicolumn{3}{c}{GTSRB} & \multicolumn{3}{c}{CelebA} \\
	  \hline
	  \multirow{2}[0]{*}{Aspect$\rightarrow$} & \multicolumn{2}{c}{Effectiveness} & Distortion & \multicolumn{2}{c}{Effectiveness} & Distortion \\
			& BA (\%) & ASR (\%) & $L_1$-norm & BA (\%) & ASR (\%) & $L_1$-norm \\
	\hline
	  Standard Training & 98.06	& $\backslash$ 	& $\backslash$	&79.70	& $\backslash$	& $\backslash$ \\
	  BadNets & 98.07 &	100 &	0.1954 &	79.06 &	100 &0.2020\\
	  ISSBA &  98.04&	99.98&	4.9572&	79.10&	99.88&	5.8129\\
	  Ours-one-target &97.61&	99.87&	0.0073&	79.17&	99.96&	0.0217\\
	  Ours-all-targets & 97.61&	99.79&	0.0076&	79.17&	99.99&	0.0213\\
	\bottomrule
	  \end{tabular}%
	\caption{Experimental results for attack effectiveness. BA and ASR mean the accuray of benign images and attack success rate, respectively.}
	\label{tab2}%
  \end{table*}%

\section{Experiment Results}
\subsection{Experimental Setup}

\paragraph{Datasets.} We employ ResNet-18 \cite{he2016deep} as the classifier, which is widely used in previous studies \cite{li2020backdoorinvisible}. We adopt two different datasets including GTSRB \cite{houben2013detection} and CelebA \cite{liu2015faceattributes}. GTSRB is a traffic signal recognition dataset with 43 categories. CelebA dataset contains  40 independent binary attribute labels. We follow the configuration proposed by previous studies \cite{NguyenT21} and choose the top three most balanced attributes including Smiling, Mouth Slightly Open, and Heavy Makeup. These attributes are concatenated to create eight classification categories. All images are resized into 128$\times$128. The number of training samples and test samples are 39209 and 12630 for GTSRB, and 162084 and 40515 for CelebA, respectively.

\paragraph{Baseline Selection.} We compare our attack with BadNets \cite{gu2017badnets} and ISSBA \cite{li2020backdoorinvisible}. BadNets is a well-known backdoor attack and is usually set as a baseline in previous studies. We employ a colourful square (6$\times$6) as the trigger in BadNets. ISSBA is a state-of-the-art invisible backdoor attack, which evades various defences. The trigger of ISSBA is generated by the official implementation released on Github. The backdoor rate of baselines is set as 0.1. We choose the label “0” as the target label for BadNets and ISSBA. Note that our approach can generate $N$ triggers for each label simultaneously (named as Ours-all-targets). We also show the result of choosing the label “0” as the target label (named as Ours-one-target).

\paragraph{Training Details.} The batch size and learning rate are set as 16 and 1e-3, respectively. The hyperparameter $\alpha$ and $\beta$ are set as 0.3 and 0.1, respectively. We keep $\alpha$ unchanged during the training process. We multiply $\beta$ by 2 every 20 epochs. The total epochs are 110 and 50 for GTSRB and CelebA, respectively.  We adopt Adam optimizer and all experiments are conducted with Pytorch 1.10 version with an NVIDIA RTX3090.
\begin{figure}[b]
	\centering
	\includegraphics[width=3.2in,clip,trim=220 240 270 160]{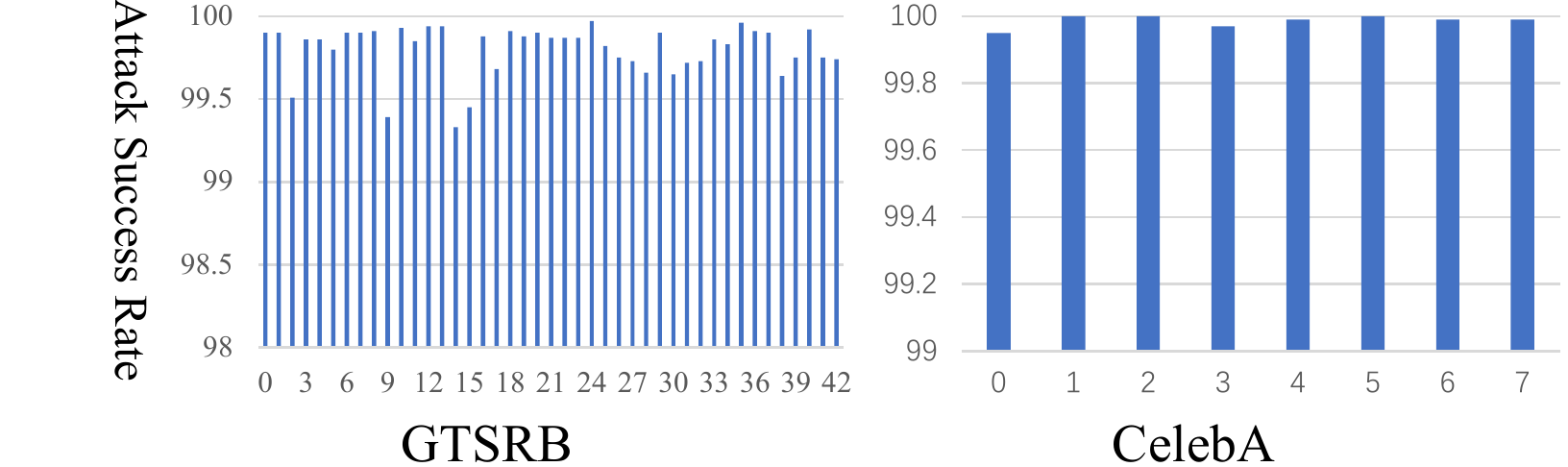}
	\caption{Attack success rate for each label.}\label{fig_3}
\end{figure}

\subsection{Attack Effectiveness and Visualization}
\begin{figure}[t]
	\centering
	\includegraphics[width=3.2in,clip,trim=320 220 320 210]{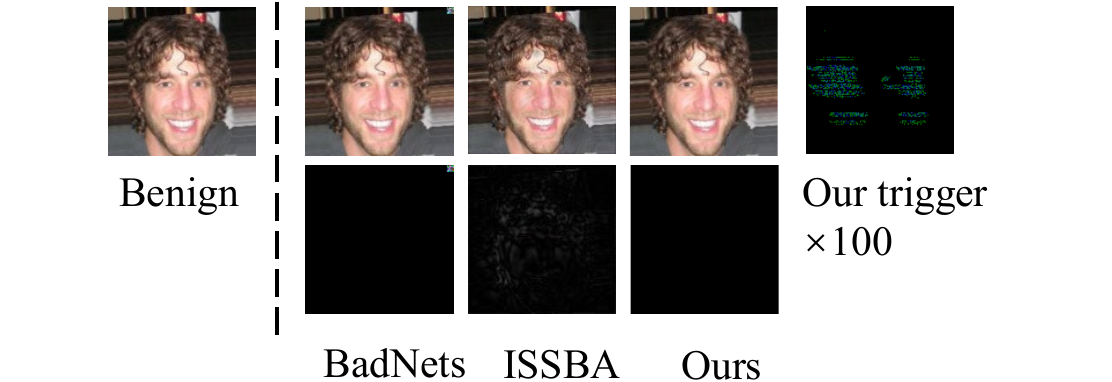}
	\caption{Visual comparison of various attacks.}\label{fig_4}
\end{figure}

For classification tasks, we employ accuracy (test set) as the metric to measure the performance of the compromised classifier. We find that all approaches achieve similar attack effectiveness in Table \ref{tab2}. Actually, most existing backdoor attacks are very similar in the aspect of attack effectiveness. The ASR is very close to 100\% and the accuracy degradation of benign images is less than 1\%. Fig. \ref{fig_3} depicts the concrete ASR over each label, and ASR is more than 98\% in most cases.
In terms of image distortion, our image distortion $L_1$-norm is significantly smaller than baselines. $L_1$-norm is calculated by $sum(abs(x_{benign}-x_{malicious}))/(channel\times height \times width)$. The $L_1$-norm is equivalent to the number of modified pixels since the maximum modification magnitude is only 1 in our approach.  Attack effectiveness of ours almost achieves perfect results. Furthermore, we visualize the trigger of ours and two baselines in Fig. \ref{fig_4}. We find that our approach achieves the best visual quality. There is a conspicuous colourful square in the top right corner of the malicious image of BadNets.

To further illustrate the modification of the trigger, we show the histogram of the trigger in Table \ref{tab3}. As depicted in Table \ref{tab3}, the number of changed pixels and magnitude is significantly smaller than the baselines. Such a small modification contributes that our approach can evade state-of-the-art trigger detection. Although the number of changed pixels of BadNets is very small, the magnitude of changed pixels is much larger than ours. Note that the practical value of modification magnitude 1,2 and 3 of BadNets is slightly larger than 0\% (around 0.0016\%). The reason is that there exists a very small proportion of pixels whose original value equals $\pm$1,$\pm$2, or $\pm$3 of the trigger. We omit these results for conciseness.

\begin{table}[b]
	\centering
	
	  \begin{tabular}{ccccccc}
	\toprule
	  MM    & 0     & 1     & 2     & 3     & 4     & $>=$5 \\
	\hline
	  BadNets & 99.78 & 0     & 0     & 0     & 0     & 0.21 \\
	  ISSBA & 9.25  & 16.83 & 14.29 & 11.56 & 9.16  & 38.89 \\
	  Ours  & 99.27 & 0.73  & 0     & 0     & 0     & 0 \\
	\hline
	  BadNets & 99.78 & 0     & 0     & 0     & 0     & 0.22 \\
	  ISSBA & 7.83  & 14.43 & 12.66 & 10.7  & 8.9   & 45.45 \\
	  Ours  & 97.87 & 2.13  & 0     & 0     & 0     & 0 \\
	\bottomrule
	  \end{tabular}%
	\caption{The histogram of the modification magnitude of the triggers. MM means the modification magnitude. The values in the table mean the proportion (\%) of corresponding modification magnitude in the total number of changed pixels. The top three rows and the bottom three rows mean the datasets of GTSRB and CelebA, respectively.}
	\label{tab3}%
  \end{table}%

\subsection{Defences}
\begin{table}[htbp]
	\centering
	  \begin{tabular}{ccc}
		\toprule
	  Dataset & Attack & Acc(\%) \\
	  \hline
	  \multirow{3}[0]{*}{GTSRB} & BadNets & 96.03 \\
			& ISSBA & 94.76 \\
			& Ours  & 49.73 \\
		\hline
	  \multirow{3}[0]{*}{CelebA} & BadNets & 99.74 \\
			& ISSBA & 85.34 \\
			& Ours  & 49.9 \\
		\bottomrule
	  \end{tabular}%
	  \caption{Detection accuray of FTD against various attacks.}
	\label{tab4}%
  \end{table}%
In this part, we evaluate the resistance of our approaches against multiple backdoor defences. First, We employ state-of-the-art trigger detection FTD \cite{zeng2021rethinking} to scan above attacks. The details of FTD have been introduced in related work. Table \ref{tab4} shows the results of FTD against various backdoor attacks. We can see that FTD is ineffective to our approach but easily detects the other two baselines. The accuracy of FTD against ours is only around 50\% which is equivalent to the random guess for a binary classification task. We only alter less than 1\% pixels with $\pm$1 modification magnitude. The trace of the trigger is too small to be detected by FTD.

\begin{figure}[t]
	\centering
	\includegraphics[width=3.2in,clip,trim=210 255 210 130]{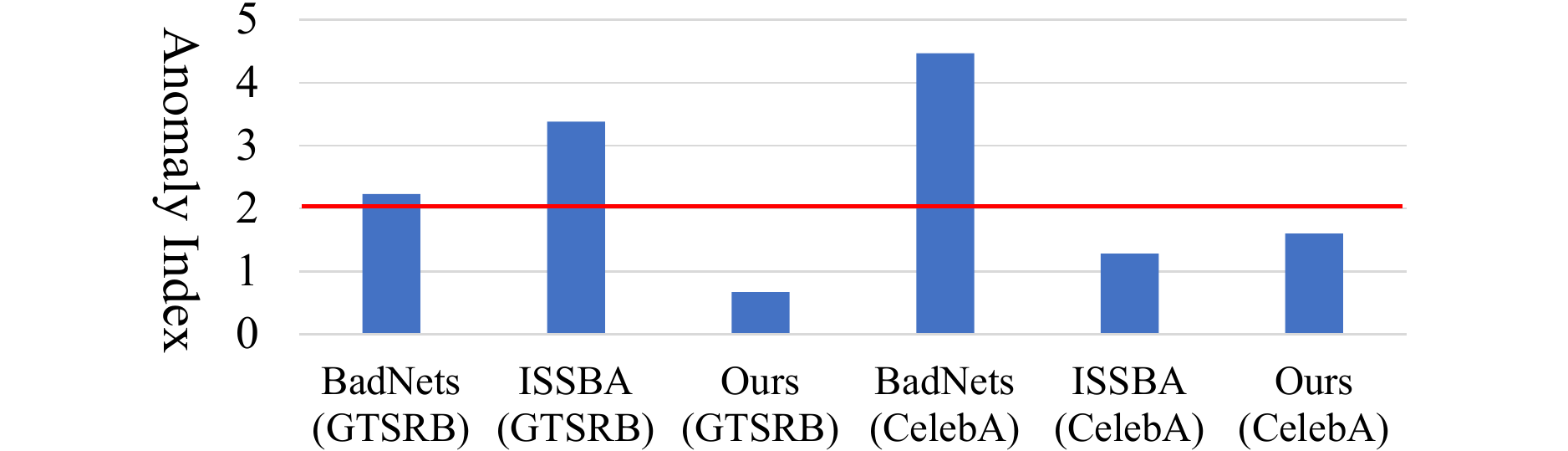}
	\caption{The experimental results of Neural Cleanse. }\label{fig_5}
\end{figure}

Apart from trigger detection, we also employ two model diagnose-based defences: Neural Cleanse \cite{wang2019neural} and Network Pruning \cite{liu2018fine}. Neural Cleanse returns an anomaly index for the suspicious classifier. If the anomaly index is more than 2, the classifier is seen as a compromised classifier. Our approach can bypass Neural Cleanse as shown in Fig. \ref{fig_5}. 
\begin{figure}[htbp]
	\centering
	\includegraphics[width=3.2in,clip,trim=240 45 310 35]{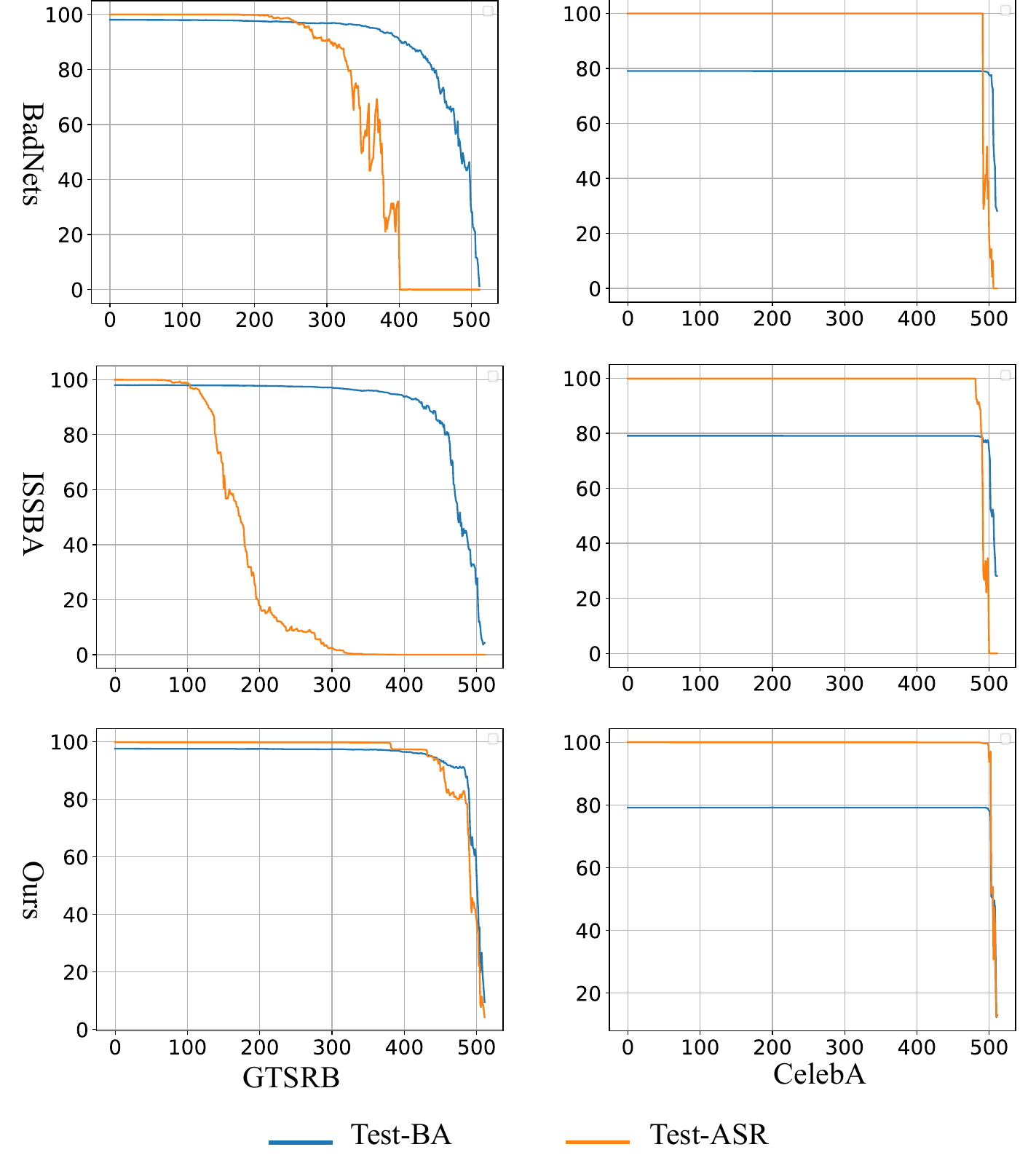}
	\caption{The experimental results of Network Pruning. }\label{fig_6}
\end{figure}
Then, we evaluate our attack against Network Pruning. Thanks to the entanglement regularization, our attack performs more resistant than baselines under two different datasets. Network Pruning cannot alleviate the backdoor without decreasing the accuracy of benign accuracy. The experimental results shown in Fig. \ref{fig_6} demonstrate that the accuracy of malicious images is entangled with benign ones. Horizatontal axis means the number of cut neurons.

\subsection{Ablation Studies} 
\begin{figure}[t]
	\centering
	\includegraphics[width=3.2in,clip,trim=265 230 350 162]{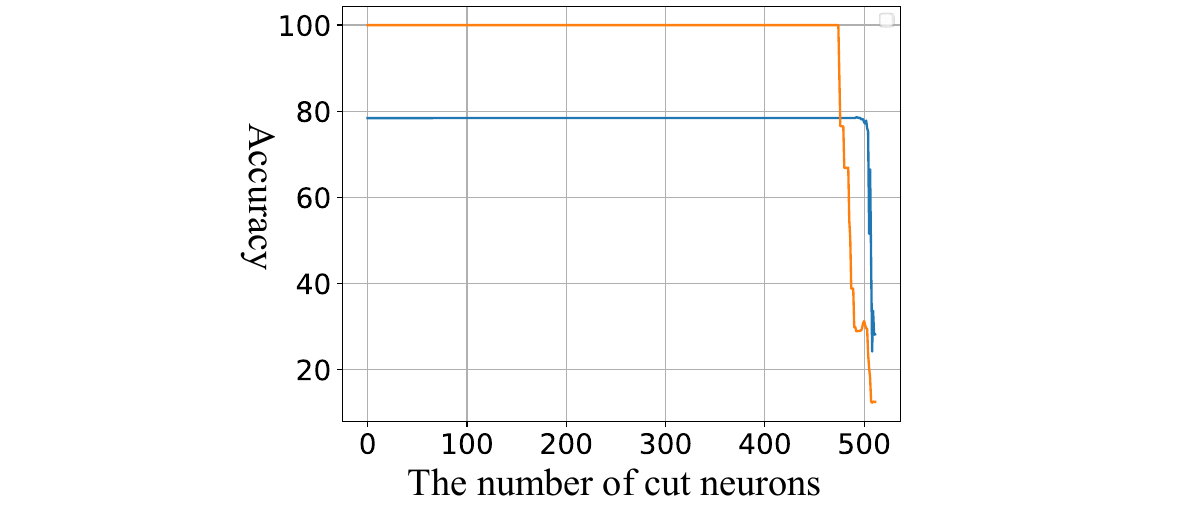}
	\caption{The experimental results of removing entanglement regularization. (CelebA). }\label{fig_7}
\end{figure}
In this part, we investigate the impacts of hyperparameter $\alpha$ and $\beta$. 
We first set $\alpha$ as 0, i.e., remove the entanglement regularization. Then, we conduct the network Pruning on the compromised model without the entanglement regularization. The experiments are shown in Fig. \ref{fig_7}. We find that the accuracy of malicious and benign can be separated like baselines. Specifically, when we cut 504 neurons, the accuracy of benign images only drops by around 3\%, whereas the accuracy of malicious images only 23.46\% (drops by around 80\%).  We also conduct experiments with large  $\alpha$ ($\alpha$=0.5 or 1). The results are similar to $\alpha$=0.3.

In terms of hyperparameter $\beta$, it minimizes the number of changed pixels. We initialize $\beta$ as 0.1 in previous experiments. We conduct experiments with a large $\beta$ (more than 0.2) and find that the cross-entropy for malicious images cannot converge. 
When we set $\beta$ as 0, the proportion of changed pixels is close to 50\%. FTD also identifies our attack with more than 90\% accuracy.

\section{Conclusions}
In this paper, we propose a novel imperceptible backdoor attack. We analyze the stealthiness of backdoor attacks from input space to feature representation. We elaborate the trigger through the sampling from a multinomial distribution which contains three probabilities +1, -1 and 0. Thanks to the elaborated trigger, we achieve both visual and statistical invisibility. In terms of the feature representation, we design the entanglement regularization to make sure the feature representations of malicious and benign images are inseparable. Extensive experiments demonstrate the effectiveness and stealthiness of our approach.

\section*{Acknowledgements}
This work was supported by the National Natural Science Foundation of China (U20B2051, U1936214).

\bibliographystyle{named}
\bibliography{ijcai22}
 
\end{document}